\RequirePackage{lineno}
\documentclass[aps,prb,showpacs, groupedaddress,twocolumn,superscriptaddress]{revtex4}
\usepackage{amsmath,graphicx,latexsym,times,color}
\usepackage{setspace}
\usepackage{hyperref}
\usepackage{array}
\usepackage{titlesec}
\usepackage{physics}

\begin{document}

\title{Highly conductive and complete spin filtering of nickel atomic contacts in a nitrogen atmosphere}

\author{Dongzhe Li}
\email{dongzhe.li@uni-konstanz.de}
\affiliation{Department of Physics, University of Konstanz, 78457 Konstanz, Germany}

\date{\today}

\begin{abstract}
    Generating efficient and highly spin-polarized currents through nanoscale junctions is essential in the field of nanoelectronics and spintronics. In this paper, using $ab$ $initio$ electron transport calculations, we predict highly conductive and perfect spin filtering of nickel atomic contacts in a nitrogen environment, where a single N$_2$ molecule sits in parallel (energetically most favorable) between two nickel electrodes. Such a particular performance is due to the wave function orthogonality between majority spin $s$-like states of ferromagnetic electrodes and the lowest unoccupied molecular orbital of the N$_2$ molecule, and thus, majority spin electrons are completely blocked at the interface. For the minority spin, on the contrary, two almost saturated conducting channels were formed due to the effective coupling between $d_{zx,zy}$ of the Ni atom and $p_{x,y}$ of the N atom, resulting in large conductance of about 1$G_0$ ($=2e^2/h$). As a consequence, a single N$_2$ molecule acts as a highly conductive and half-metallic conductor. On the other hand, the CO and NO incorporated molecular junctions exhibit rather low conductance with a partially spin-polarized current. 
    
\end{abstract}

\renewcommand{\vec}[1]{\mathbf{#1}}

\maketitle

\section{ Introduction }
\label{Intro}

A crucial issue remaining in the development of molecular spintronics \cite{sanvito-2011,Cinchetti-2017} is the manipulation of charge transport and in particular its degree of spin polarization (SP), which can be defined as $\text{SP}=(G_{\downarrow}-G_{\uparrow})/(G_{\uparrow}+G_{\downarrow}) \times 100\%$, where $G_{\uparrow}$ and $G_{\downarrow}$ are spin up (majority) and spin down (minority) conductance, respectively. Recently, it was shown that tunable SP through a single molecule can be obtained by a mechanical strain \cite{tang2015strain}, anchoring groups \cite{qiu2019enhancement}, the spin-dependent quantum interference effect \cite{Dongzhe-QI2019} etc. Unexpected large magneto-resistive ratios were observed in gas-stabilized platinum nanocontacts \cite{cespedes2014}. Control and manipulation of spin filtering in atomic and molecular junctions in order to suggest and design possible molecular-based devices with large SP and high conductance are one of the most important issues in this field. 

In ferromagnetic nanocontacts, quite generally, the electrical current is dominated by weakly polarized $s$ orbitals for both spin channels, with a smaller contribution from partially polarized minority spin $d$ orbitals, leading to limited spin-polarized current in the related nanodevices. \cite{Jacob-2005, Viljas2008, Vardimon-2016}. For instance, the SP was found to be about only 33\% in Ni atomic contacts \cite{Alex-2006}.

In molecular junctions, unlike metallic contacts, the transport between two electrodes is often mediated by a weakly coupled molecule. In particular, the molecular orbitals are expected to preserve their own symmetry and localized nature, and thus, they can be expected to exhibit properties that cannot be observed in the pure atomic contacts or even the molecule itself. For example, M. Kiguchi $et$ $al$ \cite{Kiguchi-2008} demonstrated that the electronic conductance of the Pt/benzene/Pt molecular junction is close to that of a metal nanocontact although free benzene is actually an insulator. Large SP was obtained at various spinterfaces due to spin-dependent hybridization of molecular orbitals with magnetic substrate
states \cite{Atodiresei2010,Nuala2013,Dongzhe-C60-2016,Arnoux2019}. 

Currently, charge transport of metal atomic contacts in the presence of a gas such as H$_2$, O$_2$, CO or N$_2$ has attracted a lot of interest. In the case of cobalt and palladium nanocontacts, while atomic chains were not formed for clean metals, the atomic chains can be formed in the presence of hydrogen \cite{Kiguchi2010,nakazumi2010}. High conductance as large as 1$G_0$ was observed in a single-molecule Pt/H$_2$/Pt bridge \cite{Djukic2005}. Here, $G_0=2e^2/h$ is the conductance quantum. Additionally, high spin polarization in Ni/O/Ni atomic conductors was predicted by first-principles calculations \cite{Jacob2006,Rocha2007} and confirmed later on by shot noise measurements \cite{Vardimon2015}. Moreover, it was shown that N$_2$ atmosphere greatly helps to stabilize the formation of longer Cu atomic wires, confirmed by both $ab$ $initio$ calculations \cite{Amorim2010} and mechanically controllable break junction (MCBJ) measurement \cite{Kaneko2015}. Using the MCBJ technique, the Pt/N$_2$/Pt junction was formed where the N$_2$ molecule lies parallel to current, leading to a conductance of 1$G_0$ which is close to the metal atomic contact \cite{kaneko2013electronic}.

We recently suggested a mechanism \cite{smogunov2015,Dongzhe-spinfiltering-2016}, based on robust symmetry considerations (with respect to the atomic geometry), which can lead to the upper limit of spin polarization (SP = 100\%) in $\pi$-conjugated molecular junctions. Due to symmetry match/mismatch between molecular orbitals (involved in charge transport) and ferromagnetic leads, the electron transport is carried only through minority spin-states while the $s$-channels for both spins are fully reflected at the molecule-metal interface. Here, by $ab$ $initio$ transport calculations, we predict highly conductive and perfect spin filtering of Ni nanocontacts in a nitrogen atmosphere based on similar orbital symmetry argument. From total energy calculations within density functional theory (DFT), we identified a ``parallel" configuration (i.e., single N$_2$ is placed between two Ni atoms) which is energetically most stable. Interestingly, the majority spin channel is completely blocked due to wavefunction orthogonality between $s$-states of Ni and the lowest unoccupied molecular orbital (LUMO) of the N$_2$ molecule, resulting in 100\% spin-polarized current with the junction thus playing the role of a half-metallic conductor. For the minority spin channel, on the contrary, the conductance of the single N$_2$ molecular junction is found to be as large as 1$G_0$ ($=2e^2/h$) due to effective orbital matching between $d^{\downarrow}_{zx,zy}$ and LUMO of the N$_2$ molecule, leading to a conductance is close to the pure Ni atomic contacts.

\begin{figure}[htbp]
	\centering
	\includegraphics[scale=0.45]{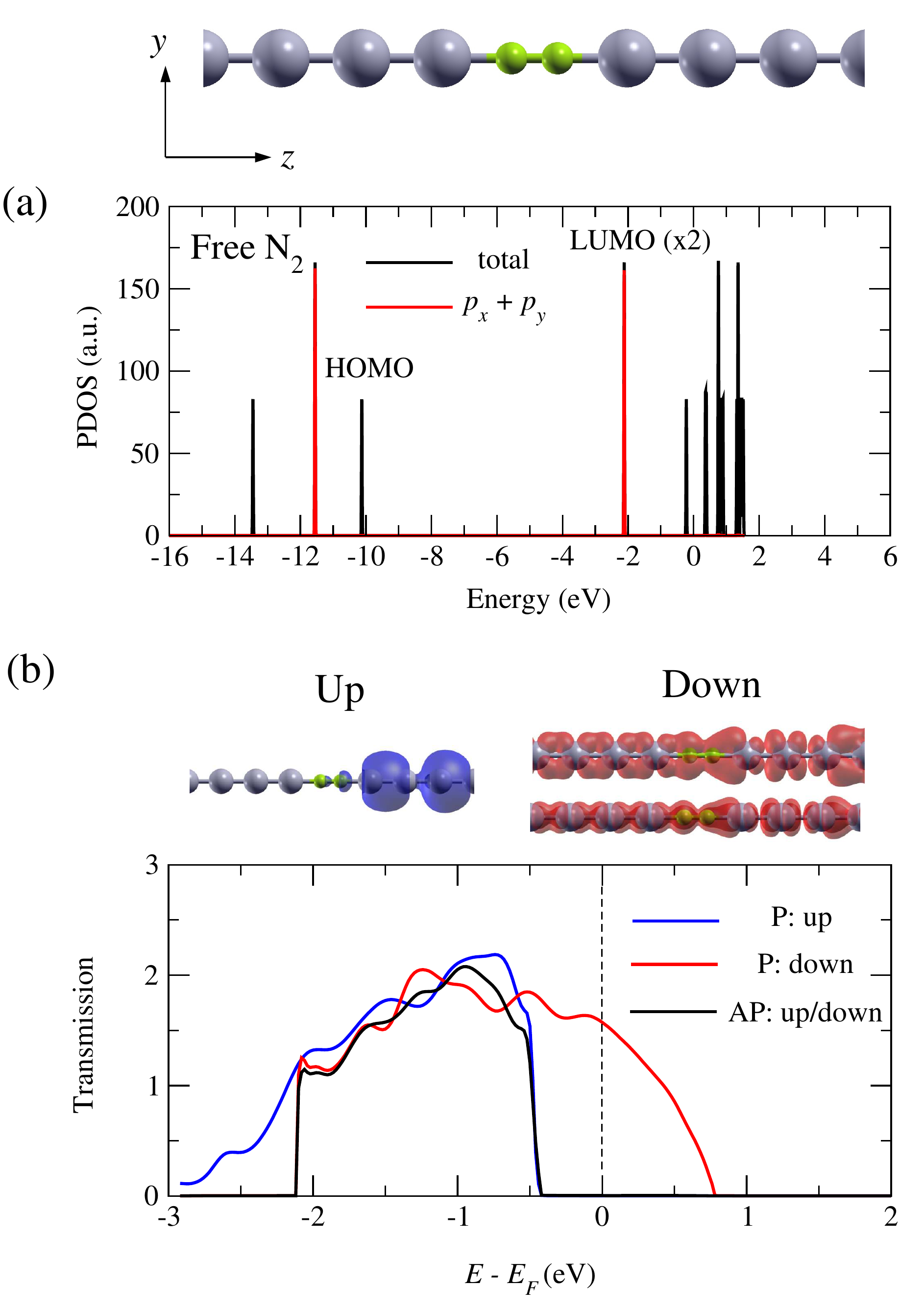}
	\caption{\label{ni-1D}
		Model molecular junction: a single N$_2$ molecule is sandwiched between two semi-infinite Ni monoatomic chains. (a) Density of states (DOS) of the N$_2$ molecule in gas phase, note that the HOMO is mainly from the $p_z$ orbital and the two-fold degenerate LUMO is from $p_x$ and $p_y$ orbitals which are orthogonal to $s$-like channels of the Ni nanowire. (b) Spin-polarized transmission functions for parallel (P) and anti-parallel (AP) spin configurations. The spin up state is completely blocked while the spin down state is partially transmitted, resulting in perfect spin filtering (SP = 100\%) and infinite magneto-resistance. Transmission eigenchannel scattering states at the Fermi energy are shown for spin up (in blue) and spin down (in red) channels.}
\end{figure}

\section{ Method}
\label{method}

Molecular junctions were described in a supercell containing a single N$_2$ molecule and two four-atom Ni pyramids attached to Ni(111) slabs containing six and six atomic layers on the left and right sides. A 4 $\times$ 4 periodicity was used in the $XY$ plane (16 atoms per layer). The atomic relaxation was carried out using plane waves $\textsc{Quantum Espresso}$ (QE) package \cite{Giannozzi2009} within DFT framework at the PBE level \cite{PBE-1996}. The same computational parameters as in Ref.~\onlinecite{Dongzhe-spinfiltering-2016} were used.

Then, $ab$ $initio$ spin-polarized electronic transport properties were evaluated using the $\textsc{Transiesta}$ code \cite{soler2002siesta,Brandbyge-2002} which employs a non-equilibrium Green's function (NEGF) formalism combined with DFT. We used Troullier-Martins norm-conserving pseudopotentials \cite{Troullier-1993}, PBE functional and an energy cutoff for the real-space mesh of 250 Ry. A double $\zeta$ plus polarization (DZP) basis set with an energy shift of 50 meV was used, which resulted, as we have checked, in a good agreement with QE results (see Fig. \ref{QE-siesta} in Appendix A). A 6 $\times$ 6 $\times$ 1 $k$-point mesh was found to be enough to obtain well converged transmission functions.

In the linear response regime, the total conductance ($G$) of magnetic systems is given by the Landauer-B\"uttiker formula,

\begin{equation}\label{conduc_eq}
G=(\frac{1}{2})G_0\sum_{\sigma}T_{\sigma}(E_F),
\end{equation}
where $G_0=2e^2/h$ is the conductance quantum ($e$ being the electron charge and $h$ Plank's constant) and $T_{\sigma}(E_F)$ is the transmission function for spin $\sigma= \uparrow, \downarrow$ at the Fermi energy. The spin-dependent conductance is then defined as $G_{\sigma}=(\frac{1}{2})G_0T_{\sigma}(E_F)$, such that $G=G_{\uparrow}+G_{\downarrow}$. Note that all the conductance values presented in this paper for both non-magnetic and magnetic systems are in units of $G_0=2e^2/h$ for the convenience of comparison.

Spin-resolved transmission function is evaluated using the NEGF formalism:

\begin{equation}
T_{\sigma}(E)=\mathrm{Tr}[\Gamma_{L,\sigma}(E)G_{\sigma}^r(E)\Gamma_{R,\sigma}(E)G_{\sigma}^a(E)],
\end{equation}

where $G_{\sigma}^{{r,a}}$ are the retarded and advanced Green's functions of the scattering region (molecule plus some parts of left and right electrodes), 

\begin{equation}
G_{\sigma}^{r/a}=[(E\pm i\eta)S - H_{\sigma}^{C} - \Sigma_{L,\sigma}^{r/a} - \Sigma_{R,\sigma}^{r/a}]^{-1}
\end{equation}
where $\eta$ is an infinitesimal positive number, $S$ is the overlap matrix, $H_{\sigma}^C$ is the Hamiltonian matrix for the scattering region and $\Sigma_{L/R,\sigma}^{r/a}$ are retarded or advanced self-energies due to left/right electrodes. Coupling matrices $\Gamma_{L/R,\sigma}$ are 
evaluated from the imaginary parts of the corresponding self-energies as $\Gamma_{L/R,\sigma} = i (\Sigma_{L/R,\sigma}^r-\Sigma_{L/R,\sigma}^a)$.

\section{Results and discussion}
\label{Results and discussion}

\subsection{Model system: Ni chain/N$_2$ junctions}
\label{model-junc}

\begin{figure}[htbp]
	\centering
	\includegraphics[scale=0.52]{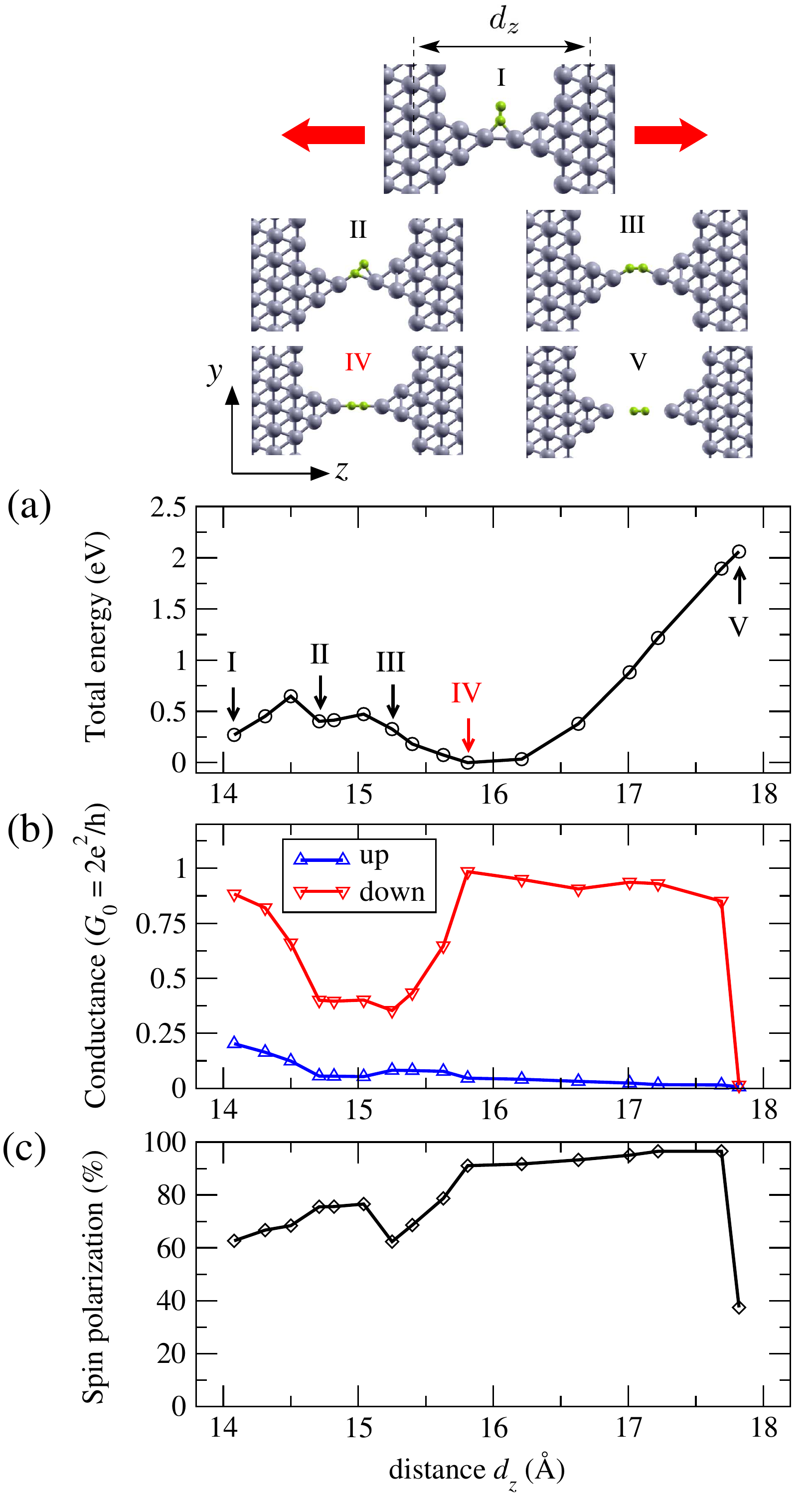}
	\caption{\label{cond-SP}
		Realistic molecular junction: N$_2$ molecular junctions with fcc-Ni(111) crystalline electrodes. (a) Total energy variation as a function of the stretch distance $d_z$ which is defined as the distance between the (111) surfaces; note that the lowest energy point is set as zero. Spin-dependent conductance (b) and the spin polarization (c) as a function of $d_z$. Five representative molecular junction geometries of the different stages of the evolution of the junction during the stretching process are presented in the top panel.
	}
\end{figure}

First we consider a simple model system in which one N$_2$ molecule sandwiched in the ``parallel" configuration (with respect to the transport direction) by two semi-infinite Ni monoatomic wires. In gas phase, as plotted in Fig. \ref{ni-1D}(a), we found that the highest occupied molecular orbital (HOMO) is mainly from $p_z$ orbital ($\bra{p_z}\ket{s} \neq 0$) while the two-fold degenerate LUMO originates from $p_{x,y}$ (plotted in red) atomic orbitals which have zero overlap with $s$-states of Ni. As described in Ref.~\onlinecite{smogunov2015}, for the Ni monoatomic chain, only one $s$ band crosses $E_F$ for spin up while five more $d$ bands are available for spin down. We plot in Fig. \ref{ni-1D}(b) transmission functions for parallel (P) and antiparallel (AP) magnetic alignments of Ni chains. Interestingly, $T(E)$ is dominated by the LUMO level and the effect of HOMO on transport is negligible due to the large band gap ($>$ 8 eV at the DFT level) for the free N$_2$ molecule. 

In the case of the P configuration, near $E_F$, the spin up transmission is essentially zero, while the spin down one shows a broad structure with the maximum transmission coefficient going up to about 2. The physical reason can be explained by orbital symmetry argument. Since the LUMO originates from $p_{x,y}$ orbitals which are orthogonal to the $s$-symmetry bands of the Ni nanowire, the spin up electron (plotted in blue) is completely blocked at the Ni-N interface. On the other hand, for the spin down (plotted in red) case electrons from only $d_{zx,zy}$ (starting at about $E_F+0.7$eV with negative band dispersion) Ni channels are transmitted due to good orbital matching with $p_{x,y}$ orbitals. This also can be clearly seen from the transmission eigenchannel scattering states at the $E_F$, shown in Fig. \ref{ni-1D}(b). As a result, we obtain $G_\uparrow$ = 0.00$G_0$ ($=2e^2/h$) and $G_\downarrow$ = 0.75$G_0$ for the P configuration, here, a single N$_2$ molecule plays the role of a perfect spin filtering component.

In the AP configuration, the emergence of almost perfect half-metallicity leads to a complete suppression of the current. This is due to the fact that the transmitted $d^{\downarrow}_{zx,zy}$ electrons from the left Ni$-$N interface are then blocked at the right Ni$-$N interface. Therefore, such junctions provide not only perfect spin filtering but also an infinite magneto-resistance (MR), which is defined as the change in electrical conductance between P and AP magnetic orientations of two ferromagnetic electrodes.

\subsection{Realistic system: Ni(111)/N$_2$ junctions}
\label{real-junc-N2}

Encouraged by the favorable results from the model system, we constructed more realistic junctions with fcc-Ni(111) electrodes for detailed analysis.

To simulate the realistic experimental situations in MCBJ, we gradually stretch the junction up to the breaking point, starting from the N$_2$ molecule which is in a upright bridge configuration (the bottom N atom is symmetrically bonded to two Ni apex atoms), shown in Fig. \ref{cond-SP} of geometry I. In the stretching process, by increasing the electrode separation $d_z$, we optimize the junction geometry by fixing three and three bottom layers on the left and right sides at every step until the junction breaks, then calculate the corresponding conductance trace. The five representative geometries depicted on top of Fig. \ref{cond-SP} are examples of the stable configurations obtained from this procedure. We note that the junctions created theoretically in the work are the idea model systems, which could be studied in MCBJ experiments. 

Our $ab~initio$ geometry optimization has shown that upon increasing the stretching distance of $d_z$ the N$_2$ configuration with Ni electrodes changes dramatically. Electronic and transport properties are summarized in Fig. \ref{cond-SP}, where we show the total energy (relative to the lowest energy configuration of $d_z = 15.81$\AA), the spin-dependent conductance [evaluated by Eq. \ref{conduc_eq} in the unit of $G_0=2e^2/h$] and the SP as a function of the electrode displacement of $d_z$.

\begin{figure}[htbp]
	\centering
	\includegraphics[scale=0.45]{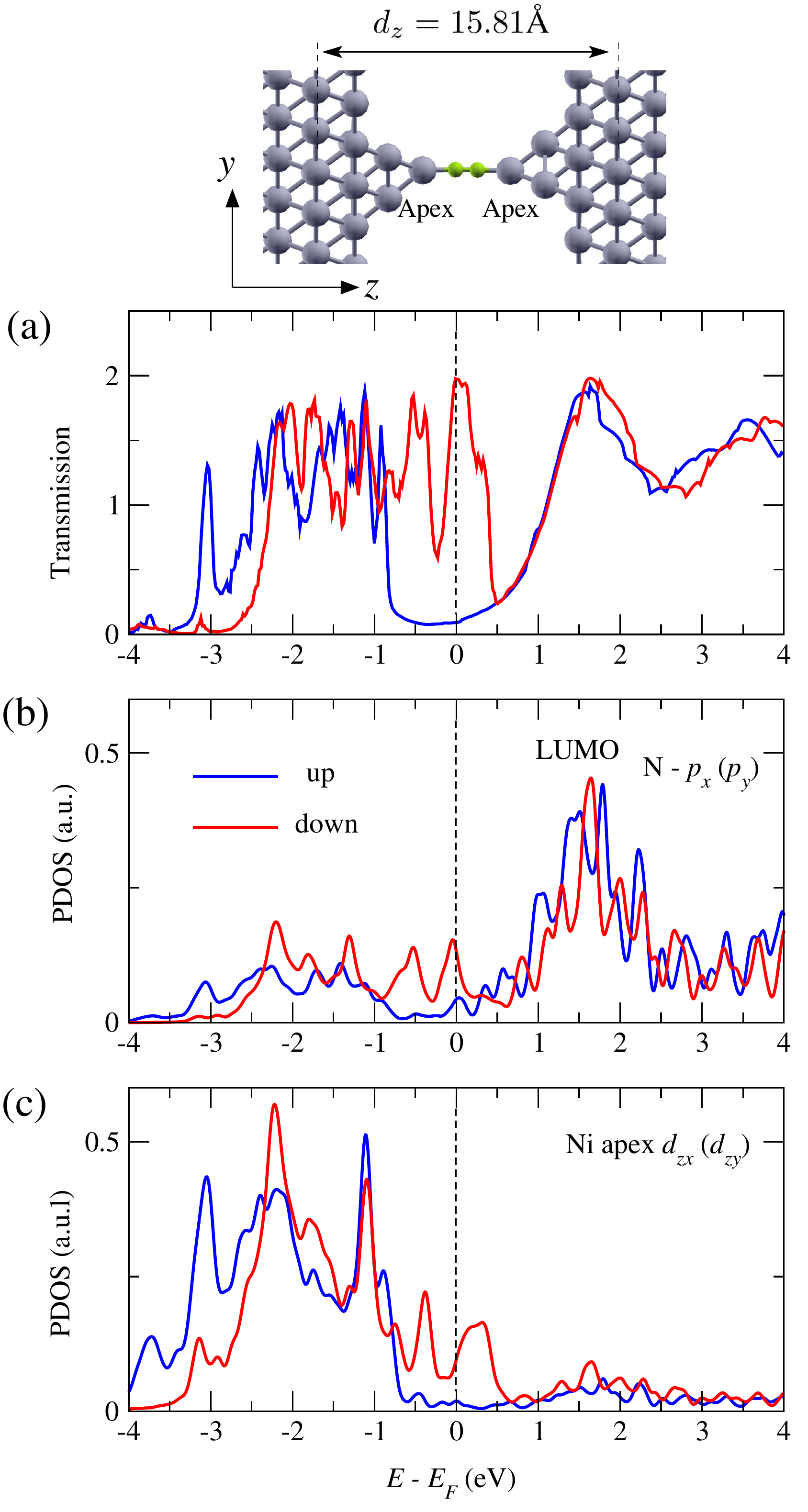}
	\caption{\label{Transmission} Electronic and transport properties in the Ni(111)/N$_2$/Ni(111) junction for the lowest energy configuration (see configuration IV in Fig. \ref{cond-SP}). (a) Spin-polarized transmission spectra for the parallel spin configuration of two Ni electrodes. (b) Spin-dependent PDOS on both nitride atoms of $p_{x,y}$ orbitals. (c) Spin-dependent PDOS on $d_{zx,zy}$ orbitals of the Ni apex atom. Note that spin up and down channels are plotted in blue and red, respectively.}
\end{figure}

Interestingly, the calculated conductance exhibits highly strain-dependent behavior for both spin channels. In general, the conductance with respect to $d_z$ can be distinguished with five different regimes as described in the following. For the first three configurations (first elastic stage), the conductance for both spins decreases monotonically. This can easily be understood by the fact that the N$_2$ molecule in the upright configuration remains stable but the hybridization strength between Ni apex atoms with the molecule decreases due to increasing $d_z$. As a result, the conductance decreases from 0.23$G_0$/0.90$G_0$ to 0.14$G_0$/0.65$G_0$ for the spin up/down channel. As the lead-lead distance is increased between 14.6 \AA~ and 15.2 \AA~the molecule begins to tilt away from the direction $y$ (second elastic stage), and the conductance for both spins further decreases leading to the low conductance plateaus of about 0.05$G_0$ and 0.37$G_0$ for spin up and down, respectively. As the contact is stretched further (15.3 \AA~$<$ $d_z$~$<$ 16.0 \AA, third elastic stage), the molecule tilts more and sits in parallel with respect to $z$ (geometry III) until forming the lowest energy configuration (geometry IV) where the molecule is placed between Ni electrodes in parallel with almost the same $x$ and $y$ coordinates as Ni apex atoms. These results are similar to those reported previously for Pt/N$_2$/Pt \cite{kaneko2013electronic} and Ag/CO/Ag \cite{balogh2014} nanojunctions. Importantly, the spin down conductance is dramatically increased up to about 1$G_0$. On the contrary, the conductance for spin down slightly decreases, resulting in a highly polarized spin current (SP $>$ 92\%). After further elongation (16.0 \AA~$<$ $d_z$~$<$ 17.4 \AA, forth elastic stage), the molecule geometry remains always in the parallel configuration showing a good mechanical stability, and thus, a clear high conductance plateau of about 0.95$G_0$ is observed for spin down while the spin up conductance is completely quenched showing almost 100\% spin-polarized current. When $d_z$ is larger than 17.6 \AA~(fifth elastic stage) the Ni$-$N bond breaks causing an abrupt decay of $T(E_F)$ almost to zero for both spin channels. Note that during the stretching evolution the N$_2$ intra-molecular bonds do not change significantly. For more detailed spin-dependent transmission functions for five representative configurations, see Appendix B.

Let us now focus on the lowest energy configuration (see geometry IV in Fig. \ref{cond-SP}) where the N$_2$ molecule sits in parallel between Ni electrodes. We plot in Fig. \ref{Transmission}(a) the spin-resolved energy dependent transmission function, where the energy is measured with respect to the Fermi energy. Interestingly, a remarkable difference of $T(E)$ curves for the two spin channels was observed. In the vicinity of $E_F$, for spin down electrons, we found a significant and broad transmission peak located between $-0.5$ eV and $0.5$ eV with respect to $E_F$ showing its maximum at $E_F$. In contrast, the transmission coefficient in the spin-up channel is strongly suppressed near $E_F$. 

To get more insights on the $T(E)$ curve, we plot in Fig. \ref{Transmission}b-c the projected density of states (PDOS) on $p_{x,y}$ of N$_2$ molecule and $d_{zx,zy}$ of the Ni apex atom, which are of major importance based on the symmetry argument discussed before in the model system. As seen in Fig. \ref{Transmission}(b), the charge transport is dominated by LUMO located at about 1.6 eV above $E_F$, resulting in a broad resonance peak in the corresponding $T(E)$. The large broadening of LUMO indicates strong hybridization between N-$p_{x,y}$ and Ni-$d_{zx,zy}$ due to almost perfect orbital matching. Moreover, near $E_F$, the PDOS of N-$p_{x,y}$ is largely dominated by spin down states. The induced spin moment on N$_2$ was found to be about -0.14 $\mu_{B}$ (``antiferromagnetic'' coupling). An important quantity to look at is the PDOS at the Ni apex atom since it provides the information on the possible nature of incoming conductance channels (see Fig. \ref{Transmission}c). Clearly, the PDOS of the Ni apex atom is dominated by the minority spin states near $E_F$ as expected in 3$d$ ferromagnetic electrodes. As a consequence, the electron transport is almost dominated by minority spin electrons; we find $G_{\uparrow} = 0.04G_0$ and $G_{\downarrow} = 0.98 G_0$, for spin up and down channels, respectively. Moreover, an orbital eigenchannel analysis reveals that the transport is mainly due to two efficient (almost saturated conductance of about 0.50$G_0$ for each channel) transmitting minority-spin channels composed of Ni-$d_{zx,zy}$ and N-$p_{x,y}$ orbitals. As a result, a single N$_2$ molecule acts as a highly conductive ($G_{\text{total}} \approx 1G_0$) and an almost 100\% spin-polarized conductor. Note that we also double-checked the conductance calculations by using the $\textsc{Pwcond}$ \cite{Alexander2004} code based on scattering theory within plane-wave basis sets as implemented in the QE package. The calculated conductance is found to be about 0.03$G_0$ and 0.79$G_0$ which is in good general agreement with $\textsc{Transiesta}$ results (see Appendix A for more details).

\begin{figure}[htbp]
	\centering
	\includegraphics[scale=0.45]{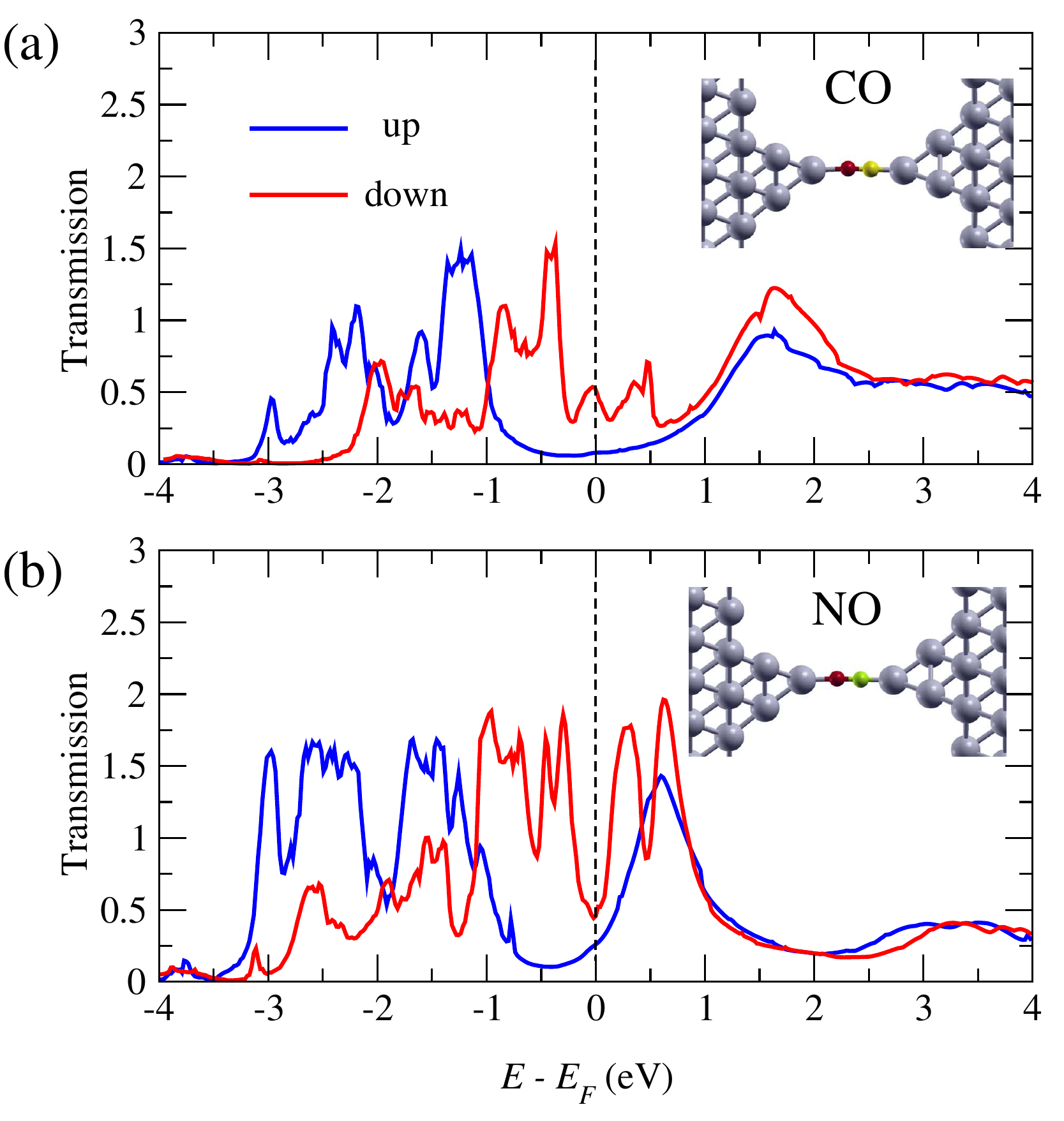}
	\caption{\label{CO-NO} Spin-dependent transmission functions for Ni/CO/Ni (a) and Ni/NO/Ni (b) junctions where the molecules sit in parallel between two electrodes. }
\end{figure}

We note that the work presented here has much better performance in terms of conductance and SP compared to the previously reported magnetic metal atomic junctions with various absorbed molecules. For example, the Ni junction bridged by a H$_2$ molecule has a conductance of approximately $0.35G_0\sim0.50G_0$ and is almost not spin-polarized (SP = 0.2\%), as confirmed by both theory \cite{li2017low} and experiment \cite{Manohar2015}. A large conductance of 1$G_0$ is reported by MCBJ experiment in Co/H$_2$/Co \cite{Untiedt2004,nakazumi2010}, but the electrical current is only partially spin-polarized. In addition, a smaller conductance of about 0.5$G_0$ is found in Ni/CO/Ni \cite{kiguchi2007effect} in the break junction experiment. Moreover, in the case of Fe/O$_2$/Fe, rather low conductance and SP of about 0.13$G_0$ and 50\% are found \cite{zheng2015}. Very recently, R. Vardimon $et$ $al$ \cite{Vardimon2015} indicated up to 100\% spin-polarized currents in the Ni/O/Ni atomic junction with measured conductance of about $0.25G_0\sim0.50G_0$

\subsection{Realistic system: Ni(111)/CO and Ni(111)/NO junctions}
\label{real-junc-NOCO}

Motivated by promising results with the N$_2$ molecule, we also investigated the transport properties of Ni atomic contacts in the presence of CO and NO molecules in the parallel configuration (also fully relaxed). These molecules are chosen because they have electronic properties in gas phase very similar to those of N$_2$, namely, LUMO is the linear combination of $p_{x,y}$ atomic orbitals while HOMO is from $p_{z}$ and $s$ orbitals. These molecular junctions have been created with various metals by MCBJ \cite{kiguchi2007effect} or via scanning tunneling microscope (STM) with functionalized tips \cite{Martina2015,Akitoshi2018}. Note that the Ni$-$CO contact spin polarizes and the major current is carried by the minority spin channel (red line), while the majority spin channel (blue line) only has a minor contribution at the Fermi level, shown in Fig. \ref{CO-NO}(a). Clearly, the coupling between $d_{zx,zy}$ of apex Ni atoms and $p_{x,y}$ of CO molecule determines the transmission. The calculated conductance is about 0.32$G_0$ with SP of about 75\% which is in general agreement with a previously reported experimental result for the Ni/CO/Ni junction \cite{kiguchi2007effect}. In the case of the NO molecule (see Fig. \ref{CO-NO}b), the LUMO position is about 1.0 eV closer to $E_F$ compared to the CO and N$_2$ molecules. On the contrary, the broadening of the LUMO is smaller in the NO than in the other two molecules. For spin down, two peaks at about 0.25 eV and 0.75 eV just above $E_F$ originate from the spin-split hybridized $d$-states of Ni apex atoms. The calculated conductance was found to be about 0.13$G_0$ and 0.23$G_0$ for majority and minority spins, respectively, resulting in a rather small SP of about 28\%.

Finally, let us discuss the generality of symmetry arguments and accuracy issues. The symmetry reasoning presented here in principle work for a rather broad class of atomic and molecular junctions, where no $s$-like channels are available in the molecular spin-valve set-up. They could be realized by MCBJ or STM measurements, for instance, atomic chains such as carbon sandwiched by magnetic electrodes \cite{smogunov2015}, $\pi$-conjugated molecular spin valves such as polythiophene \cite{Dongzhe-spinfiltering-2016} and the atomic scale contacts in the presence of gas molecules such as N$_2$. In this work, we base our theoretical prediction on the NEGF-DFT technique which is widely recognized as a robust approach, providing an understanding of the behavior on a good qualitative basis at a reasonable computational cost, as seen, for example, in Ref.~\onlinecite{Kiguchi-2008, kaneko2013electronic} for nanoscale junctions. However, from quantitative point of view, standard DFT could underestimate the HOMO-LUMO gap and overestimate the conductance. $GW$ self-energy \cite{Strange-2011} has been recognized as a good approximation to describe accurately the energy level alignment at the molecule-metal interface but much more computationally demanding. The DFT error is in general systematic, thus conductance ratios are usually in good agreement with experiment, for instance the rectification ratios predicted by NEGF-DFT were found to be reliable \cite{elbing2005single, Batra2013}. So we believe that the spin filtering ratios predicted in this work are reliable as well. In summary, the DFT-error introduced here plays a role (for the overestimation of conductance values) in a quantitative basis but should not affect significantly our main conclusions resulting from robust symmetry mismatch arguments.

\begin{figure}[htbp]
	\centering
	\includegraphics[scale=0.45]{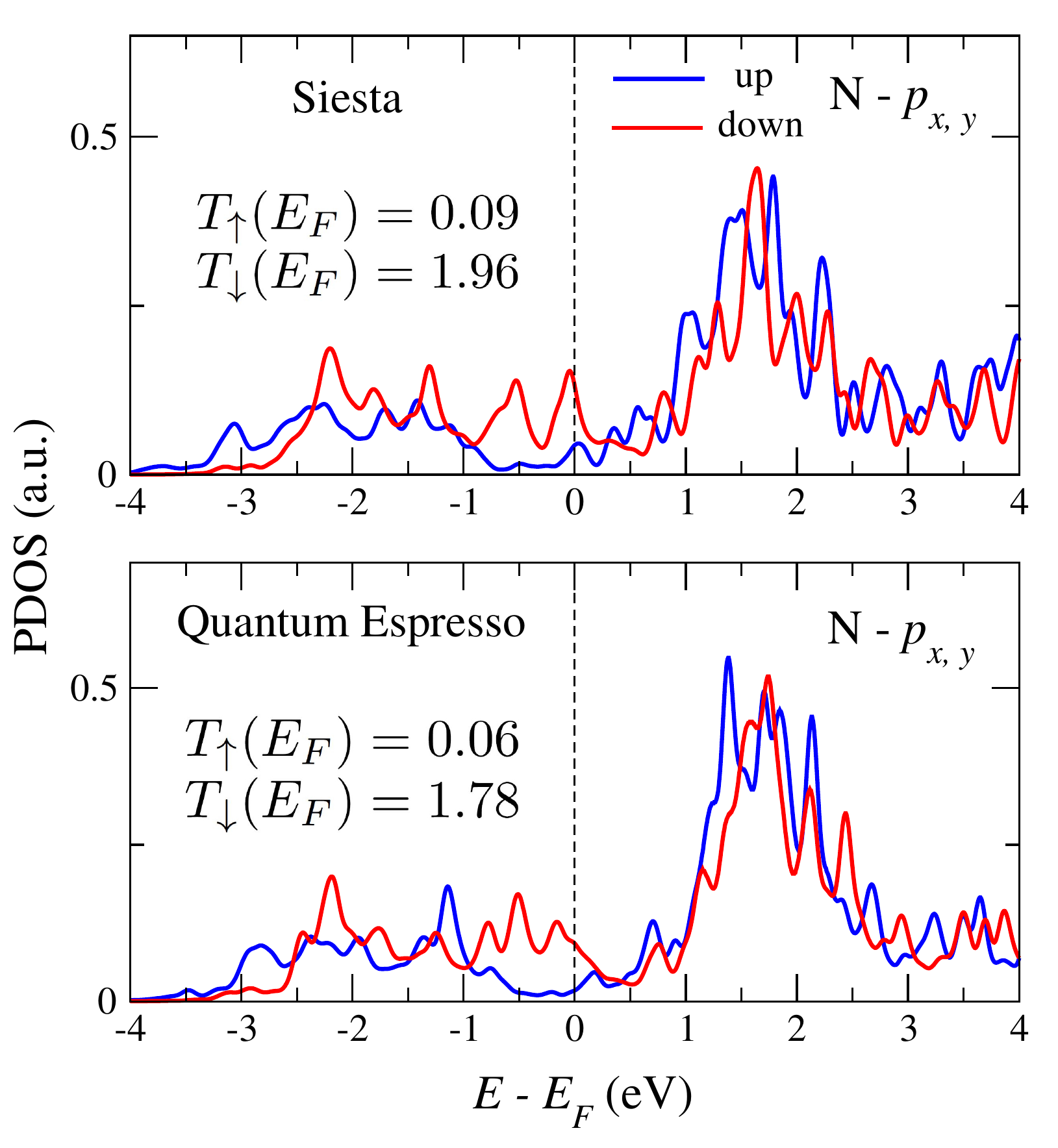}
	\caption{\label{QE-siesta} Comparison between localized atomic-orbital basis $\textsc{Siesta}$ and plane-wave basis $\textsc{QE}$ results. Calculated spin-dependent PDOS on nitride $p_{x,y}$ by $\textsc{Siesta}$ (top) and $\textsc{QE}$ (bottom). The transmission coefficient at $E_F$ evaluated by $\textsc{Transisesta}$ and $\textsc{Pwcond}$ were also shown. A good general agreement was found between two different codes.}
\end{figure}

\section{ Conclusions}
\label{conclusions}

To conclude, using a combination of density functional theory and non-equilibrium Green's function method, we predict highly conductive and perfect spin filtering of nickel magnetic point contacts in a nitrogen atmosphere. For Ni/N$_2$/Ni nanocontacts we identified that an energetically stable configuration is the N$_2$ molecule providing a parallel configuration (with respect to the current flow) between two Ni apex atoms. The N$_2$ junction shows a high conductance of about 1$G_0$ ($=2e^2/h$), which is comparable to that of corresponding pure Ni atomic contacts. More importantly, the majority spin conductance is almost quenched near $E_F$ due to the symmetry mismatch between $s$-states of the Ni and $p_{x,y}$ of the N$_2$ molecule. In contrast, for the minority spin, two almost saturated conducting channels originate from the effective coupling between $p_{x,y}$-symmetry of the N$_2$ molecule and $d_{zx,zy}$-symmetry of the Ni apex atoms due to the symmetry. However, if the N$_2$ is replaced by CO or NO, the conductance, as well as the SP, is reduced due to less pronounced spin-split hybridized molecular states near $E_F$. We hope that our theoretical prediction may inspire further experimental explorations (i.e. measurements of shot noise) to reach the upper limit of spin polarization with large conductance in magnetic metal nanocontacts with various absorbed molecules.

{\bf Acknowledgments}: D.L. wants to thank S. Lamowski for helpful comments. D.L. was supported by the Alexander von Humboldt Foundation through a
Fellowship for Postdoctoral Researchers.
\\

\section*{Appendix A: Comparison between $\textsc{siesta}$ and $\textsc{QE}$ results}

In order to check the reliability of the DZP basis sets used in this work, we compared the electronic and transport properties in the Ni/N$_2$/Ni junction (geometry IV presented in Fig. \ref{Transmission}) using the $\textsc{siesta}$ \cite{soler2002siesta} and $\textsc{quantum espresso (QE)}$\cite{Giannozzi2009} packages. Within the plane-wave basis sets, $ab$ $initio$ transport properties were evaluated using the $\textsc{pwcond}$ code \cite{Alexander2004}, which is part of the QE package. Separate calculations were performed for the leads (complex band structure) and scattering regions, which were combined using the wave-function matching technique. 

In Fig. \ref{QE-siesta} we plot the spin-resolved PDOS on $p_{x,y}$ of N atoms using $\textsc{Sisesta}$ (top) and $\textsc{QE}$ (bottom). Good general agreement was found in terms of energy level alignment, namely, LUMO is located about 1.6 eV above $E_F$ and the PDOS at $E_F$ is largely dominated by minority spin. The induced spin moment was found to be about -0.14$\mu_B$ and -0.12$\mu_B$ for $\textsc{Sisesta}$ and $\textsc{QE}$, respectively. The conductance calculated by $\textsc{pwcond}$ was about 0.03$G_0$ and 0.79$G_0$ for spin up and down which is slightly smaller than the corresponding $\textsc{Transisesta}$ results.

\begin{figure}[htbp]
	\centering
	\includegraphics[scale=0.45]{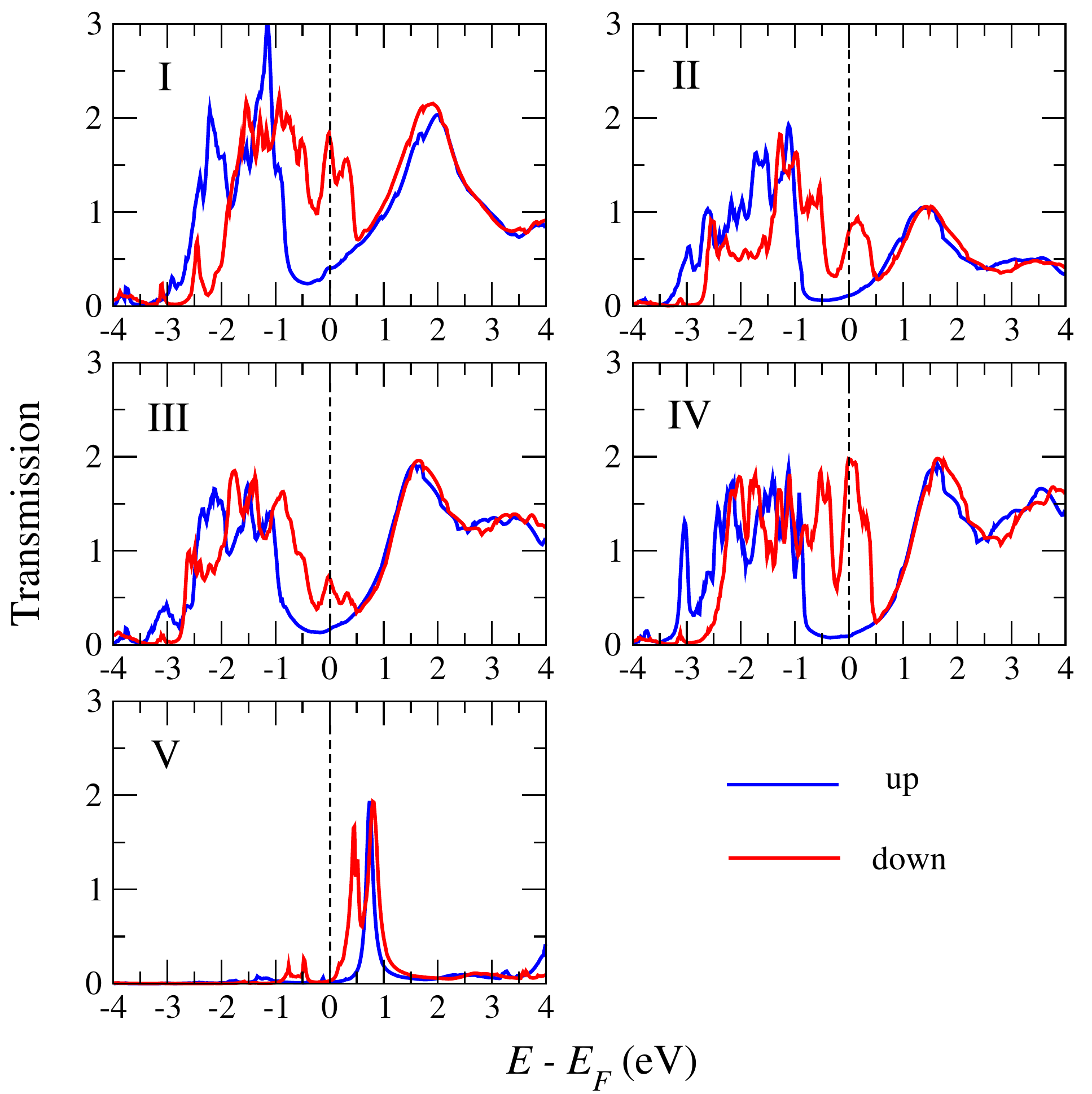}
	\caption{\label{T-complete} Spin-dependent transmission functions of five representative relaxed geometries with different elongation distances presented in Fig. \ref{cond-SP}. Blue and red lines denote spin up and down channels, respectively.}
\end{figure}

\section*{Appendix B: Spin-dependent $\textit{T(E)}$ for five representative geometries}

We display in Fig. \ref{T-complete} the spin-polarized transmission spectra for stretched Ni/N$_2$/Ni junctions for five representative geometries at equilibrium as presented in Fig. \ref{Transmission}. During the stretching process, the positions and structures of the transmission peaks are nearly invariant but the magnitudes are changed dramatically near $E_F$ when the junction is stretched. In the case of configuration I, the non-negligible spin up transmission coefficient of about 0.45 is from the partially transmitted $s$-channel of direct Ni$-$Ni contact. Note that a significant reduction of spin up transmission is observed when the junction is elongated to configuration IV. In contrast, for the spin down channel, the conductance was decreased first and then increases up to about 1$G_0$ (geometry IV). From configuration I to II, the decreases conductance is related to the breaking of the direct Ni$-$Ni bond at the nanocontact. And dramatic enhancement of spin down conductance from geometry III to IV can be explained by orbital symmetry argument and augmented molecular level amplitude at $E_F$ in geometry IV due to enhanced hybridization strength. Finally, when the junction breaks (geometry V), we observe a sharp LUMO resonance due to very weak hybridization between molecule and electrodes, and thus the conductance for both spins is almost quenched.

\bibliographystyle{apsrev}
\bibliography{References}

\end{document}